\documentstyle[aps,prl,preprint]{revtex}
\draft
\begin{document}

\title{Single-point velocity distribution in turbulence}
\author{G. Falkovich$^a$ and V. Lebedev$^{a,b}$}
\address{$^a$ Physics of Complex Systems, Weizmann Institute of Science,
Rehovot 76100, Israel
\\ $^{b}$ Landau Inst. for Theor. Physics, Moscow, Kosygina 2,
117940, Russia}
\maketitle

\begin{abstract}
We show that the tails of the single-point velocity probability distribution
function (PDF) are generally non-Gaussian in developed turbulence. By using
instanton formalism for the Navier-Stokes equation, we establish the relation
between the PDF tails of the velocity and those of the external forcing.
In particular, we show that a Gaussian random force having correlation
scale $L$ and correlation time $\tau$ produces velocity PDF tails
$\ln{\cal P}(v)\propto-v^4$ at $v\gg v_{\rm rms}, L/\tau$. For a
short-correlated forcing when $\tau\ll L/v_{\rm rms}$ there is an intermediate
asymptotics $\ln {\cal P}(v)\propto-v^3$ at  $L/\tau\gg v\gg v_{\rm rms}$.
\end{abstract}
Early experimental data on skewness and flatness of the velocity field prompted 
one
to believe that the single-point velocity PDF in developed
turbulence is generally close to Gaussian \cite{53Bat,MY}. A possible reasoning 
may be
that large-scale motions (that give the main contribution into velocity
statistics at a point) are connected to a random external forcing $f$ then
velocity $v(t)=\int_0^tf(t')dt'$ has to be Gaussian if $t$ is larger then 
the correlation time $\tau$ of the forcing, irrespective of the statistics of
$f$. That would be the case if the force was the only agent affecting
velocity. However, there are also nonlinearity (leading to instability and
break-up of large-scale motions) and viscosity (that dissipates small-scale
modes appeared as a result of the instability). Let us first explain the simple 
physics
involved and formulate the predictions following from physical arguments,
then we develop the formalism which gives the predicted PDF tails.
Qualitatively, one may describe the interplay between external force and
nonlinearity in the following way. Force $f$ pumps velocity $v\sim ft$ until
the time $t_*\sim L/v$ when nonlinearity restricts the growth. The relation
between velocity and forcing can thus be suggested as follows: $v^2\sim fL$.
Therefore, velocity's PDF can be obtained by substituting $f\sim v^2/L$ into
force's PDF ${\cal P}_f$: ${\cal P}(v)\sim{\cal P}_f(v^2/L)$.
Those arguments presume that $t_*$ is less than the correlation time $\tau$ of
forcing. If opposite is the case $t_*\gg\tau$ then the law of velocity growth 
is different $v^2\sim f^2t\tau$, so that the velocity increases up to
$v^3\sim f^2L\tau$; the short-correlated pumping is effectively
Gaussian ${\cal P}_f\sim\exp(-f^2)$ and the velocity's PDF is
${\cal P}(v)\sim\exp[-(v/v_{\rm rms})^3]$.
We see that velocity PDF is expected to be dependent on the statistics of
the force.

Actual mechanism of restriction for the Navier-Stokes equation
(instability of a large-scale flow leading to a cascade that provides for
a viscous dissipation) is irrelevant for the above arguments. What matters
is that we deal with the system of the hydrodynamic type so that nonlinear
time $t_*$ can be estimated as $L/v$. For example, the same argument goes
for Burgers equation where $t_*$ is a breaking time and viscous dissipation
at a shock prevents further growth \cite{97BFKL}. It is interesting that
viscosity does not enter above estimates yet it's existence is crucial for
the whole picture to be valid.

Let us stress  that the above arguments can be only applied to rare events
with velocity and force being much larger than their root-mean-square values
when the influence of background fluctuations can be neglected.
The above predictions are thus made for PDF tails. For a
a nonlinear dissipative system, it is generally difficult to relate
an output statistics to the statistics of the input (be it initial
conditions or external force).  Our aim here is to show that it is
possible, nevertheless, to relate the probabilities of rare fluctuations that
is to relate the tails of the PDFs of the force and the field that is forced
respectively. A rigorous way to describe rare fluctuations is the instanton
formalism recently developed for turbulence \cite{96FKLM} and employed for
obtaining PDF tails in different problems \cite{96GM,97BFKL,96Chert}.

The main idea of the method is that the tails are described by saddle-point
configurations in the path integral for the correlation functions of the
turbulent variable (say, velocity ${\bbox v}$). We call the configuration
instanton because of a finite lifetime.
One may call it also optimal fluctuation since it corresponds to the extremum
of the probability.

We start with the Navier-Stokes equation
\begin{equation}
\partial_t{v}_\alpha
+ v_\beta\nabla_\beta v_\alpha
-\nu \nabla^2 v_\alpha
+\nabla_\alpha P= f_\alpha \,,
\label{NS}\end{equation}
where ${\bbox f}$ is a random force (per unit mass) 
pumping the energy into the system
and $\nu$ is the viscosity coefficient. Incompressibility is assumed so that
${\rm div}\,{\bbox v}={\rm div}\,{\bbox f}=0$. The field $P$ in
(\ref{NS}) is the pressure divided by the mass density $\rho$. Velocity
correlation functions can be presented as path integrals which form is
determined by the statistics of pumping.  Let us first consider a Gaussian
forcing with the correlation function $\langle f_\alpha(t,{\bbox r})
f_\beta(t',{\bbox r}')\rangle= \Xi_{\alpha\beta}(t-t',{\bbox r}-{\bbox r}')$
which is assumed to decay on the scale $\tau$ as a function of the first
argument and on the scale $L$ as a function of the second one. Then moments
of the velocity can be written as path integrals:
\begin{eqnarray} &&
\langle v^{2n}\rangle=\int {\cal D}p\,{\cal D}v\,{\cal D}P\,{\cal D}Q\,
\exp\left(i{\cal I}+2n\ln{ v}\right) \,,
\label{j5} \end{eqnarray}
where ${\bbox p}$ is an auxiliary field and the effective action
has the following form  \cite{73MSR}.

\begin{eqnarray} &&
{\cal I}=\int\, dt\, d{\bbox r}\,
\bigl[p_\alpha\bigl(\partial_t v_\alpha +
v_\beta\nabla_\beta v_\alpha-\nu\nabla^2 v_\alpha
+ \nabla_\alpha P\bigr)
\nonumber\\&&
+\, Q \nabla_\alpha v_\alpha\bigr]
+{i\over2}\int dt' dt d{\bbox r}'d{\bbox r}\,
\Xi_{\alpha\beta}(t-t',{\bbox r}-{\bbox r}') p_\alpha p'_\beta \,.
\label{j4} \end{eqnarray}
The independent fields $P$ and $Q$ play the role of Lagrange multipliers
enforcing the incompressibility conditions on the velocity and the response
field: $\nabla_\alpha v_\alpha= \nabla_\alpha p_\alpha=0$.

The tails of the velocity PDF are determined by high moments with $n\gg1$
which can be found by applying the saddle-point method to the integral
(\ref{j5}): $\langle v^{2n}\rangle=u^n(0,0)\exp({\cal I}_{extr})$. 
The configuration ${\bbox u}(t,{\bbox r})$,
${\bbox p}(t,{\bbox r})$, $P(t,{\bbox r})$ and $Q(t,{\bbox r})$
corresponding to a saddle point is our instanton.
The extremum conditions for the argument of the exponent in
(\ref{j5}) determining the instanton give two dynamical equations
\begin{eqnarray}  &&
\partial_t{u}_\alpha
+ u_\beta \nabla_\beta u_\alpha
-\nu \nabla^2 u_\alpha
+\nabla_\alpha P
\nonumber \\ &&
=-i\int dt'\,d{\bbox r}'\,\Xi_{\alpha\beta}(t-t',{\bbox r}-{\bbox r}')
p_\beta(t',{\bbox r}') \,,
\label{eu1} \\  &&
\partial_t{p}_\alpha
-p_\beta\nabla_\alpha u_\beta
+ u_\beta \nabla_\beta p_\alpha
+ \nu \nabla^2 p_\alpha+\nabla_\alpha Q
\nonumber\\&&\
=2\,i\,n\,\delta(t)\delta({\bbox r})u_\alpha/u^2\,,
\label{eu2} \end{eqnarray}
and two incompressibility constraints
\begin{eqnarray} &&
\nabla^2 P= -\nabla_\alpha
(u_\beta \nabla_\beta u_\alpha) \,.
\label{j6} \\ &&
\nabla^2 Q=\nabla_\alpha(p_\beta \nabla_\alpha u_\beta
- u_\beta \nabla_\beta p_\alpha) \ .
\label{jj6} \end{eqnarray}

As it was explained in \cite{96FKLM}, the system (\ref{eu1}--\ref{jj6}) is to
be solved at negative time because it describes the prehistory that leads to
a given measured value of the velocity at $t=0$. Since $p=0$ at $t>0$ then the
right-hand side of (\ref{eu2}) directly produces the value of $p$ at $t=-0$:
\begin{equation}
p_\alpha(0,{\bbox r})=-2in\left[
\delta_{\alpha\beta}\delta({
\bbox r})-\frac{1}{4\pi r^3}
\left(\delta_{\alpha\beta}
-\frac{3r_\alpha r_\beta}{r^2}
\right)\right] \frac{u_\beta(0,0)}{u^2(0,0)} \ .
\label{ji6} \end{equation}

The equations (\ref{eu1},\ref{eu2}) are to be solved at $-\infty<t<0$ with
the constraints (\ref{j6},\ref{jj6}) under the boundary conditions (\ref{ji6})
and $u\rightarrow0$ as $t\rightarrow-\infty$. Instanton approach thus reduces
the statistical problem (finding PDF tails) to the dynamical problem of
finding a particular solution of the deterministic equations 
(\ref{eu1},\ref{eu2}).
Note that the instanton equations are deterministic
they present already substantial simplification with respect to the
initial Navier-Stokes equation with a random force
(which is not suprising because they describe only part
 of velocity statistics, namely the tails of the single-point PDF). 
Still, the system (\ref{eu1},\ref{eu2}) is quite complicated and it's complete 
solution (which will provide an important information about the spatio-temporal
domains with high velocity) is still ahead of us. The main difficulty
is an effective spatial nonlocality of the constraints (\ref{j6},\ref{jj6}).

Our goal in this paper is to show that a definite statement about the
functional form of velocity PDF tails can be obtained from the analysis of the
symmetries of the instanton equations without actually finding a solution. 
Let us show how the symmetry analysis gives $n$-dependence of $\langle
v^{2n}\rangle$ which determines the functional form of PDF tails.  What is
important to state is that both fields ${\bbox u}$ and ${\bbox p}$ decay at
moving backwards in time during some characteristic time $t_*$ which we call
the lifetime of the instanton.  Actually it is the same time which we
qualitatively discussed above at treating rare events with strong forcing
since the instanton presents just the space-time picture of those typical
events contributing to $\langle u^{2n}\rangle$.  Symmetry analysis depends on
whether the pumping correlation time $\tau$ is larger or smaller than the
instanton lifetime $t_*$. Let us first consider the case $\tau\gg t_*$, that
makes it possible to consider the pumping correlation function $\Xi$ as time
independent. In this case, the parameter $n$ can be excluded from
(\ref{eu1}--\ref{jj6}) by the rescaling transformation \begin{eqnarray} &&
t\to X^{-1}t\,, \ \  {\bbox u}\to X{\bbox u}\,, \ \ P\to X^{2}P\,, \ \  Q\to
X^{4}Q\,, \nonumber \\ && \nu\to X\nu\,, \ \ \ {\bbox p}\to X^{3}{\bbox p}
\,,\ \ \  X^4=n\ .  \label{tra} \end{eqnarray} That gives a general
$n$-dependence of the velocity \begin{equation} {\bbox u}=
n^{1/4}{\bbox\varphi}(\nu^4/n)\,, \label{tru}\end{equation} where
dimensionless function $\varphi$ is expected to go to some constant when it's
argument goes to zero. This is equivalent to the physically plausible
assumption that the high velocity moments are viscosity independent. Under
such an assumption the $n$-dependence of the instanton solution $u\propto
n^{1/4}$ gives the following $n$-dependence of the moment $\langle
v^{2n}\rangle\propto n^{n/2}$ which corresponds to the PDF tail
\begin{equation} \ln{\cal P}(v^2)\propto -v^4\,.  \label{tail1}\end{equation}
Note that the integral term in the action ${\cal I}_{extr}\propto n$ i.e. the
factor $\exp({\cal I}_{extr})$ gives only subleading contribution into
(\ref{tail1}).  If $L/\tau\lesssim v_{\rm rms}$ (where $v_{\rm rms}$ is the
typical value of the velocity fluctuations) then the asymptotics (\ref{tail1})
is realized at $v\gg v_{\rm rms}$ and $\ln{\cal P}\sim-(v/v_{\rm rms})^4$.  In
the opposite limit of a short-correlated pumping with $L/\tau\gg v_{\rm rms}$,
an intermediate asymptotics exists where the pumping correlation function
$\Xi$ can be treated as delta-correlated in time:  $\Xi(t,{\bbox
r})=\delta(t)\chi({\bbox r})$.  Then the only changes in (\ref{tra}) are
\begin{equation} {\bbox p}\to X^{2}{\bbox p}\,,\quad Q\to X^3Q\,,\quad
X^3=n\,.  \label{tra1} \end{equation} That leads to the law (cf. 
\cite{96FKLM}) \begin{equation} \ln{\cal P}(v^2)\sim -(v/v_{\rm rms})^3 \,,
\label{tail2}\end{equation} which is valid at $L/\tau\gg v\gg v_{\rm rms}$.
For larger $v$, the asymptotics (\ref{tail1}) is realized. We thus see that
for Gaussian pumping the velocity PDF always decreases faster than Gaussian.

Note that the above consideration can be readily extended for the
consideration of velocity differences 
${ w}(\bf r)=|{\bbox v}(\bf r)-{\bbox v}(0)|$ at any $r$.
There is an intermediate region $w_{rms}(r)\ll w(r)\ll v_{\rm rms}$ where
${\cal P}(w)$ is not described by the direct instanton formalism. 
The same formulas (\ref{tail1},\ref{tail2}) as well as below (\ref{l14}) 
describe the remote tails of ${\cal P}(w)$ at $w\gg v_{\rm rms}$, which are 
probably unaccessible experimentally at present.

Contrary, faster-than-Gaussian decay of a single-point velocity PDF was
recently observed both in experiments \cite{relief} and in numerical
simulations \cite{91VM,jim,KK}, the qualitative reason for that (short life
time of strong fluctuations) was discussed in \cite{KK}. Unfortunately,
except the qualitative statement on faster-than Gaussian decay there were no
definite data in \cite{91VM,jim,KK} to allow for a quantitative comparison
with (\ref{tail1},\ref{tail2}), this remains for the future work.

The transformations (\ref{tra},\ref{tra1}) shows that the lifetime of the
instanton is short for large $n$: $t_*\propto n^{-1/3}$ for the fast pumping
$\tau\ll t_*$ and $t_*\propto n^{-1/4}$ for the slow pumping $\tau\gg t_*$.
Therefore at large enough $n$ we always deal with a slow pumping. In other
words, distant tails of the velocity PDF are determined by a simultaneous
statistics of the pumping (which is not necessary Gaussian) rather than by
integral over time as one would naively expect.

As we shall show now, the same property ($\tau\gg t_*$) is correct for a
non-Gaussian statistics of the pumping. Indeed, the above procedure can be
readily generalized for an arbitrary pumping statistics when instead of
(\ref{j5}) one has \begin{eqnarray} && \langle|{\bbox v}|^{2n}\rangle= \int
{\cal D}f{\cal D}p{\cal D}v{\cal D}P{\cal D}Q {\cal
P}_f\exp\left(i\tilde{\cal I}+2n\ln{|{\bbox v}|}\right) \ .  \label{l5}
\end{eqnarray} Here, we substituted the simultaneous PDF of pumping ${\cal
P}_f(f)$ since we treat the slow pumping and the effective action
$\tilde{\cal I}$ figuring in (\ref{l5}) for the Navier-Stokes equation has
now the following form \begin{eqnarray} && \tilde{\cal I}=\int\,dt\, d{\bbox
r}\, \bigl[ p_\alpha\bigl(\partial_t v_\alpha + v_\beta\nabla_\beta
v_\alpha-\nu\nabla^2 v_\alpha + \nabla_\alpha P\bigr)\nonumber\\&& +\, Q
\nabla_\alpha v_\alpha -p_\alpha f_\alpha\bigr] \,.  \label{l4}
\end{eqnarray} The saddle-point equation for the instanton velocity ${\bbox
u}$ is now written as \begin{eqnarray} \partial_t u_\alpha +
u_\beta\nabla_\beta u_\alpha-\nu\nabla^2 u_\alpha + \nabla_\alpha P=f_\alpha
\,, \label{l6} \end{eqnarray} with the relation \begin{eqnarray} \int
dt\,p_\alpha=-i\delta\ln{\cal P}_f(f)/\delta f_\alpha\,.  \label{l8}
\end{eqnarray} The equation for ${\bbox p}$ and the incompressibility
constraints are the same (\ref{eu2},\ref{j6},\ref{jj6}).

Let us assume the tail of ${\cal M}=-\ln{\cal P}_f$
to be scale invariant with the exponent $a$:
\begin{equation}
{\cal M}(X{\bbox f})=X^a{\cal M}({\bbox f}) \,,
\label{l9} \end{equation}
Now, we can generalize (\ref{tra})
\begin{eqnarray}&&
t\to X^{-1}t\,, \ \  {\bbox u}\to X{\bbox u}\,, \ \
P\to X^{2}P\,, \nonumber \\ &&
\nu\to X\nu\,, \ \ {\bbox f}\to X^2 {\bbox f}\,,
\label{l10} \\ &&
{\bbox p}\to X^{2a-1}{\bbox p}\,, \ \
Q\to X^{2a}Q \,, \ \ X^{2a}=n \,,
\nonumber \end{eqnarray}
keeping the form of the equations (\ref{l6}), (\ref{l8}) and
removing $n$ from the equation (\ref{eu2}).
Using the law (\ref{l10}) we can find how the characteristic velocity scales
with $n$. As above the answer can be expressed in terms of the behavior of
the tail of the simultaneous PDF ${\cal P}(v^2)$ for the velocity:
\begin{equation}
\ln{\cal P}(v^2)\propto -v^{2a} \,.
\label{l14} \end{equation}
The lifetime $t_*$ of our instanton scales as $t_*\propto n^{-1/(2a)}$
that is decreases with $n$ for any positive $a$ which justifies
our consideration. For Gaussian statistics $a=2$ and we reproduce (\ref{tail1}).

Note that the velocity PDF decays always faster than that of the force; in
particular, Gaussian velocity would correspond to the force PDF decaying
exponentially.

The expression (\ref{j4}) implies homogeneous pumping. Meanwhile, the
transformations (\ref{tra},\ref{tra1},\ref{l10}) do not transform coordinates
and can be generalized for spatially ingomogeneous pumping statistics.
Therefore our conclusions are true also for a more physical case of an
inhomogeneous pumping, in particular acting on the boundaries of the flow.

To conclude, let us describe the status of the results obtained: 
Under the assumptions that the solutions of the instanton equations exist and
the probability of finding very high velocity is viscosity 
independent we found 
how that probability is related to the statistics of the force.

\acknowledgements

We are grateful to U. Frisch, R. Kraichnan, A. Noullez and E. Siggia for
useful discussions. The work was partially supported  by the Israel Science
Foundation, by a research grant from the Cemach and Anna Oiserman Research
Fund and the Minerva Center for Nonlinear Physics. V.L. is a Meyerhoff
Visiting Professor at the Weizmann Institute.

\end{document}